\documentclass[twocolumn, amsmath, amssymb, aps, prl, superscriptaddress]{revtex4-2}

\usepackage{graphicx}
\usepackage{bm}
\usepackage{xcolor}
\usepackage[colorlinks=true, allcolors=blue]{hyperref}
\usepackage{braket}

\begin{document}

\title{Trotter Scars: Trotter Error Suppression in  Quantum Simulation}

\author{Bozhen Zhou}
\thanks{zhoubozhen@itp.ac.cn}
\affiliation{Institute of Theoretical Physics, Chinese Academy of Sciences, Beijing 100190, China}

\author{Qi Zhao}
\thanks{zhaoqi@cs.hku.hk}
\affiliation{QICI Quantum Information and Computation Initiative, School of Computing and Data Science,The University of Hong Kong, Pokfulam Road, Hong Kong}

\author{Pan Zhang}
\thanks{panzhang@itp.ac.cn}
\affiliation{Institute of Theoretical Physics, Chinese Academy of Sciences, Beijing 100190, China}
\affiliation{School of Fundamental Physics and Mathematical Sciences, Hangzhou Institute for Advanced Study, UCAS, Hangzhou 310024, China}
\affiliation{Beijing Academy of Quantum Information Sciences, Beijing 100193, China}

\begin{abstract}
Recent studies have shown that Trotter errors are highly initial-state dependent and that standard upper bounds often substantially overestimate them. However, the mechanism underlying anomalously small Trotter errors and a systematic route to identifying error-resilient states remain unclear. Using interaction-picture perturbation theory, we derive an analytical expression for the leading-order Trotter error in the eigenbasis of the Hamiltonian. Our analysis shows that initial states supported on spectrally commensurate energy ladders exhibit strongly suppressed error growth together with persistent Loschmidt revivals. We refer to such states as \emph{Trotter scars}. To identify such states, we further introduce a model-agnostic variational framework. Its loss function can be built from Trotterized dynamics alone, which allows the search to reach system sizes beyond exact diagonalization. The optimized states at small sizes moreover follow regular patterns that extend to larger sizes. We demonstrate our theory in three spin models, where the optimized states exhibit the predicted persistent Loschmidt revivals and strongly suppressed error growth. We further conducted experiments on a $17$-qubit superconducting quantum processor and successfully realized the Trotter-scar states and demonstrated the Trotter error suppression in quantum simulations.
\end{abstract}

\maketitle


The Trotter--Suzuki product formula lies at the heart of digital quantum simulation~\cite{lloyd1996universal}, providing the most direct bridge from a many-body Hamiltonian to an executable gate sequence.
In universal gate-based quantum computing, the continuous time evolution $U(t)=e^{-iHt}$ generated by a Hamiltonian $H$ must be decomposed into a sequence of discrete quantum gates. The Trotter--Suzuki formula achieves this by splitting $H$ into separately exponentiable parts, yielding circuits composed solely of local unitaries.
However, this discretization inevitably introduces Trotter errors that accumulate with simulation time~\cite{suzuki1990fractal, preskill2018quantum, childs2019nearly}.
In recent years, quantum computational advantage in many-body dynamics simulation has attracted broad attention~\cite{daley2022practical}, with landmark digital-circuit experiments on superconducting processors~\cite{kim2023evidence, arute2019quantum, abanin2025constructive} all relying on Trotterized evolution as the core algorithmic primitive.
Alternative approaches, including linear combinations of unitaries~\cite{childs2012hamiltonian}, quantum signal processing~\cite{low2017optimal}, Taylor-series methods~\cite{berry2015simulating}, and hybrid analogue--digital architectures~\cite{andersen2025thermalization, liu2026prethermalization}, can offer asymptotically superior scaling. Nevertheless, the Trotter formula remains the method of choice for near-term implementations owing to its simplicity and low hardware overhead, making the understanding and suppression of Trotter errors a pressing priority~\cite{campbell2019random, childs2019faster, rendon2024improved,childs2018toward}.

Toward this goal, rigorous worst-case bounds on Trotter errors have been established with commutator scaling~\cite{childs2021theory}. However, such bounds are often overly pessimistic. Motivated by this limitation, recent theoretical advances have shifted the error-analysis paradigm from pessimistic worst-case commutator bounds toward state or observable-dependent perspectives \cite{heyl2019quantum, burgarth2023state, burgarth2024strong, zhao2025entanglement,sieberer2019digital, kargi2021quantum, chinni2021trotter,zhao2022random,yu2025observable,sahinoglu2020hamiltonian, An2021time}.
Notably, it has been established that random initial states yield an average-case performance \cite{zhao2022random,chen2023average} with a quadratic improvement in system-size scaling, and that sufficient entanglement can accelerate the simulation toward this statistical average \cite{zhao2025entanglement}. However, these results primarily describe the bulk behavior of the Hilbert space; the existence and physical origin of specific non-thermal states that exhibit error resilience far exceeding the average-case limit remain entirely unknown.

In this Letter, we connect the Trotter error to the spectral structure of the underlying Hamiltonian and identify a class of dynamical anomalies that exhibit both a suppression of Trotter‑error growth by orders of magnitude beyond average‑case expectations and long‑lived Loschmidt revivals.
Crucially, this mechanism operates identically for product formulas of any order, from first-order Lie-Trotter to arbitrary $2k$-order Suzuki formulas. We further propose a variational optimization framework based on a product-state ansatz that can automatically identify error-minimizing initial states for a given Hamiltonian. Applying this framework to the Heisenberg chain with a transverse field, the Stark spin chain, and the PXP model, where exact or approximate commensurate ladders arise from symmetry, strong fields, or special invariant subspaces, we discover a new class of states that we call Trotter scars. The naming draws an analogy with quantum many-body scars~\cite{bernien2017probing, turner2018weak, serbyn2021quantum, zhang2023manybody}, where a small set of nonthermal eigenstates embedded in an otherwise chaotic spectrum gives rise to anomalous revivals. Similarly, Trotter scars are supported on a vanishingly small fraction of the Hilbert space yet exhibit anomalously suppressed Trotter errors accompanied by long-lived periodic oscillations in the Loschmidt echo. Because Trotter scars are product states, they can be prepared by a single layer of single-qubit rotations. We implement all three models on a superconducting quantum processor and observe the predicted revivals.

{\it Perturbative analysis---}We begin by analyzing how the Trotter error of a given initial state is governed by the spectral properties of the Hamiltonian.
Consider an $n$-qubit Hamiltonian with nearest-neighbor coupling, decomposed into even- and odd-bond terms $H=H_{e}+H_{o}$, where each subset contains mutually commuting interactions.
Standard product formulas approximate the propagator $e^{-iH\Delta t}$ by splitting it into separately exponentiable parts.
The simplest instance is the first-order Lie--Trotter formula
$\mathcal{S}_1(\Delta t) = e^{-iH_o\,\Delta t}\,e^{-iH_e\,\Delta t}$.
More generally, the $(2k)$th-order Suzuki formula~\cite{suzuki1990fractal} is constructed recursively from the second-order symmetric decomposition
\begin{equation}
\mathcal{S}_2(\Delta t) = e^{-iH_o \Delta t/2}\,e^{-iH_e\Delta t}\,e^{-i H_o\Delta t/2},
\label{eq:S2}
\end{equation}
via
\begin{equation}
\mathcal{S}_{2k+2}(\Delta t)=\bigl[\mathcal{S}_{2k}(p_k\Delta t)\bigr]^2 \mathcal{S}_{2k}(s_k\Delta t)\bigl[\mathcal{S}_{2k}(p_k\Delta t)\bigr]^2,
\end{equation}
where $p_k=1/(4-4^{1/(2k+1)})$ and $s_k=1-4p_k$.
For concreteness, we illustrate the analysis with $\mathcal{S}_2$. The generalization to all orders, including $\mathcal{S}_1$, is deferred to the Supplemental Material~\cite{SM}.
Using the Baker--Campbell--Hausdorff expansion, the effective generator of this discrete evolution can be written as
\begin{equation}
H_{\mathrm{eff}} = H + \Delta t^{2} K_2 + O(\Delta t^{4}),
\end{equation}
where $K_2$ is the leading-order error kernel composed of nested commutators.
We define the error vector as the difference between the ideal and the Trotterized time-evolution of an initial state $|\psi_0\rangle$:
\begin{equation}
|\delta\psi(t)\rangle \equiv  (e^{-iHt} - e^{-iH_{\mathrm{eff}}t}) |\psi_0\rangle.
\end{equation}
Treating $V = \Delta t^2 K_2$ as a perturbation, we can use standard interaction-picture perturbation theory to express the error state at time $t$ as:
\begin{equation}
|\delta\psi(t)\rangle \approx i \Delta t^2 \int_0^t d\tau \, e^{-iH(t-\tau)} K_2 e^{-iH\tau} |\psi_0\rangle.
\label{eq:delta_psi_integral}
\end{equation}
Expanding in the eigenbasis $H|n\rangle = E_n|n\rangle$ with $|\psi_0\rangle = \sum_m c_m |m\rangle$ and evaluating the time integral (see Supplemental Material~\cite{SM} for details), the squared error norm becomes
\begin{equation}
\|\delta\psi(t)\|^2 \approx \Delta t^4 \sum_{n} \left| \sum_{m} c_m (K_2)_{nm} e^{\frac{i\omega_{nm} t}{2}} \frac{2\sin\frac{\omega_{nm} t}{2}}{\omega_{nm}} \right|^2,
\label{eq:error_norm_spectral}
\end{equation}
where $\omega_{nm} \equiv E_n - E_m$. This analytical expression reveals the crucial role of the spectrum. The stroboscopic factor $\sin(\omega_{nm} t/2)$ controls whether each spectral contribution accumulates or is periodically suppressed, while $1/\omega_{nm}$ provides an amplitude weight favoring small energy differences. Crucially, Eq.~(\ref{eq:error_norm_spectral}) suggests a mechanism for error suppression when the initial state is supported on a spectral ladder. If the relevant energy gaps are commensurate, $\omega_{nm}=k\Omega$ with $k\in\mathbb{Z}$, the spectrum defines a set of stroboscopic times $t_p=2\pi p/\Omega$ at which the stroboscopic factor vanishes for all off-diagonal terms with $k\neq 0$:
$\sin(\omega_{nm} t_p/2)=\sin(k p\pi)=0$.
Compared to the generic case where all frequency components contribute incoherently, the ladder structure causes a substantial fraction of the off-diagonal terms in Eq.~(\ref{eq:error_norm_spectral}) to be suppressed near the stroboscopic times. While diagonal contributions ($\omega_{nn}=0$) and degeneracies  ($\omega_{nm}=0$) persist, this partial cancellation leads to an overall reduction in Trotter error and characteristic temporal oscillations, rather than monotonic growth.

The most distinctive signature of such ladder-supported states appears in the Loschmidt echo,
$\mathcal{F}(t)=|\langle\psi_0|e^{-iHt}|\psi_0\rangle|^2$.
Expanding in the eigenbasis yields
$\mathcal{F}(t) = \left|\sum_n |c_n|^2 e^{-iE_n t}\right|^2$.
If the spectral weight is concentrated on an equidistant ladder $E_n\approx E_0+n\Omega$, then
\begin{equation}
\mathcal{F}(t) \approx \left|\sum_n |c_n|^2 e^{-in\Omega t}\right|^2,
\end{equation}
which is periodic with period $T=2\pi/\Omega$. At stroboscopic times $t_p=pT$, the phases realign perfectly, yielding pronounced revivals $\mathcal{F}(t_p)\approx 1$. This robust revival of the Loschmidt echo provides a clear experimental criterion for identifying initial states with ladder-supported spectral structure. The periodic Loschmidt revivals are reminiscent of quantum many-body scars, where special eigenstates forming approximately equally spaced energies give rise to persistent oscillations that resist thermalization. Trotter scars share this spectral structure. The two phenomena are nevertheless not equivalent. Trotter-error suppression requires not only spectral commensurability but also favorable matrix elements of the error kernel within the ladder subspace. Many-body scar states do not automatically satisfy the latter condition. The variational framework introduced below is precisely designed to identify states that fulfill both requirements simultaneously.
    
\begin{figure*}[t]
    \centering
    \includegraphics[width=0.9\textwidth]{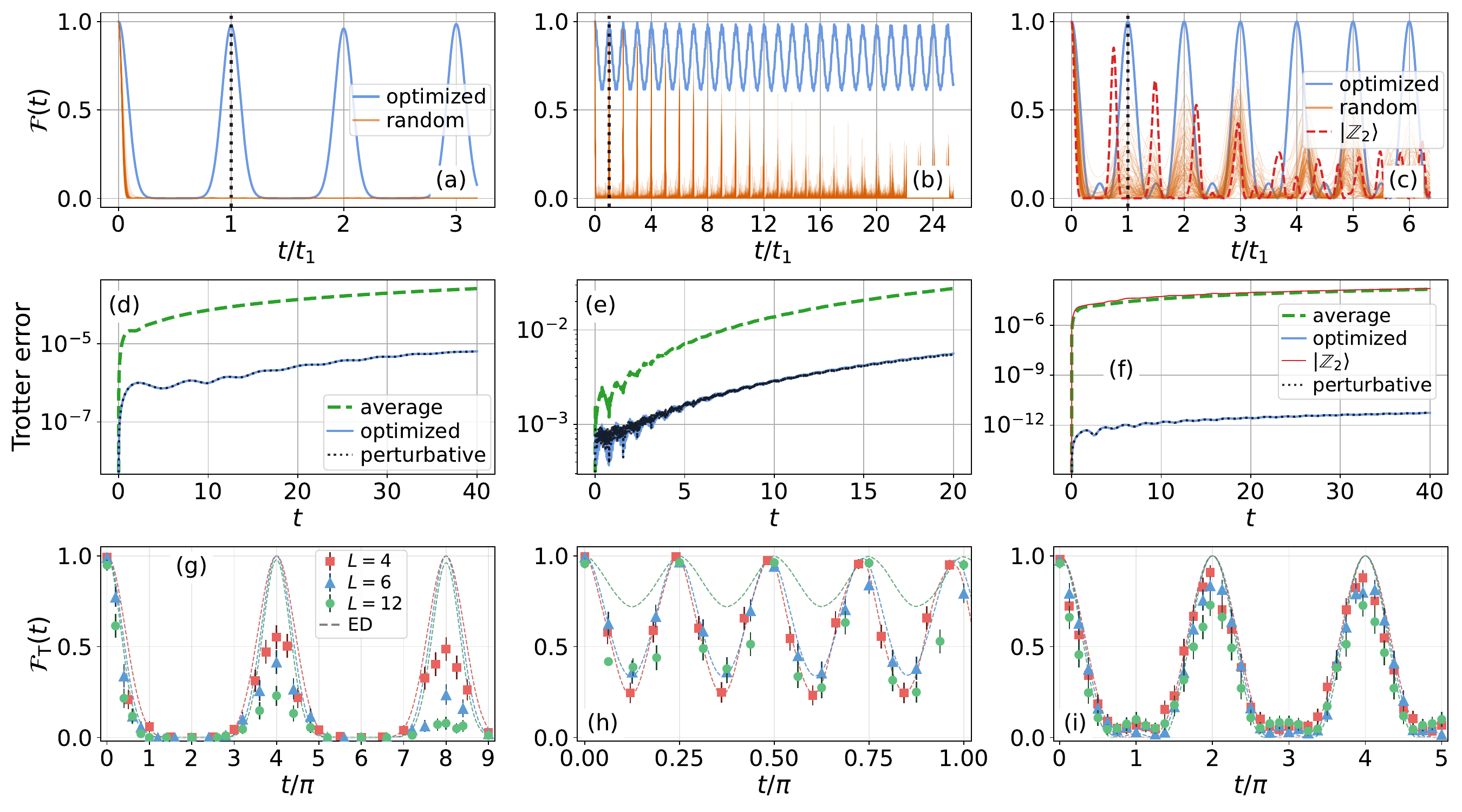}
    \caption{Characterization of Trotter scars in the Heisenberg chain (first column, $h_x=0.5$), the Stark spin chain (second column, $h_x=0.8$, $h_y=0.9$, $h_z=4.0$), and the PXP model (third column).
    (a,b,c)~Loschmidt echo $\mathcal{F}(t)$ for the optimized state and 100 random states. The vertical dashed line marks the stroboscopic time $t_1=2\pi/\Omega$.
    (d,e,f)~Accumulated Trotter error $\|\delta\psi(t)\|$ for the optimized state and the random-state average over 100 initial states. The black dotted lines are the perturbative result obtained from Eq.~\eqref{eq:error_norm_spectral}, which agrees well with the exact Trotter error. Results in (a$\sim$f) are for $L=12$ with optimization time $T_l=10$ and Trotter step $\Delta t=0.01$.
    (g,h,i)~Trotterized Loschmidt echo of the optimized Trotter-scar states measured on a superconducting processor. Markers are return probabilities obtained from 512 shots per circuit for different system sizes. For visibility the error bars denote three times the statistical uncertainty from finite sampling. Dashed lines are exact diagonalization of the corresponding full Hamiltonian in continuous time. Circuits realize the second-order product formula in (g) and (h), and the first-order formula in (i). The stroboscopic time $t_1$ is $4\pi$ for the Heisenberg chain, $\pi/4$ for the Stark chain and $2\pi$ for the PXP model.}
    \label{fig:trotter_error}
\end{figure*}

{\it Variational method---} The spectral analysis above predicts which states are error-resilient but does not prescribe how to find them in a given model. We now introduce a model-agnostic variational framework that automatically identifies initial states minimizing the Trotter error over a finite time window. Given the parametrized initial state
$|\psi(\boldsymbol{\theta}, \boldsymbol{\phi})\rangle
=\bigotimes_{j=1}^{L}
U_j(\theta_j,\phi_j)\,|0\rangle_j$,
where $U_j(\theta_j,\phi_j)=R_y(\phi_j)\,R_x(\theta_j)$ is a single-qubit rotation with two variational parameters per site, we minimize the composite loss function
\begin{equation}
\mathcal{L}(\boldsymbol{\theta}, \boldsymbol{\phi})
=
l_1 \left\|\delta\psi(T_l) \right\|+ \frac{l_2}{T_l} \int_0^{T_l} \mathcal{F}_{\mathrm{T}}(t)\, dt,
\end{equation}
which combines the accumulated Trotter error at time $T_l$ with the Trotterized Loschmidt echo $\mathcal{F}_{\mathrm{T}}(t) \equiv |\langle \psi(0) | \Psi_{\mathrm{Trotter}}(t) \rangle|^2$, i.e.\ the overlap between the initial state and the Trotterized time-evolved state, averaged over the optimization window $[0, T_l]$.
The second term penalizes states that remain nearly stationary under Trotterized evolution and thereby prevents convergence to trivial fixed points (see End Matter for the optimization details). To further reach system sizes beyond exact diagonalization, the first term can be replaced by the norm of the difference of two Trotterized trajectories. The optimized states at small sizes moreover reveal structure that yields deterministic constructions at larger sizes~\cite{SM}.
To validate the perturbative predictions and demonstrate the effectiveness of the variational framework, we apply it to the following spin systems whose spectra contain sectors or subspaces that support the commensurate ladder condition.

{\it Heisenberg chain---}We consider an isotropic Heisenberg spin-$\tfrac12$ chain in a uniform transverse field,
\begin{equation}
H = H_{\mathrm{iso}} + h_x \sum_j S_j^x ,
\label{eq:Heisenberg_model}
\end{equation}
with $S_j^{x,y,z}=\sigma_j^{x,y,z}/2$ and
\begin{equation}
H_{\mathrm{iso}} =
\sum_{j=1}^{L-1}
\left(
S_j^x S_{j+1}^x
+ S_j^y S_{j+1}^y
+ S_j^z S_{j+1}^z
\right),
\end{equation}
where the exchange coupling is set to unity. The SU(2) symmetry of $H_{\mathrm{iso}}$ organizes the spectrum into degenerate multiplets labeled by total spin $S$. The transverse field splits each multiplet into an exactly equidistant ladder with spacing $\Omega=h_x$ (see End Matter).

{\it Stark spin chain---}We next consider a spin-$\tfrac12$ chain of length $L$ subject to a linear Stark potential, where the ladder structure is approximate rather than symmetry-exact,
\begin{equation}
H =
J_x \sum_{j=1}^{L-1} \sigma_j^x \sigma_{j+1}^x
+ \sum_{j=1}^{L}
\left(
h_x \sigma_j^x
+ h_y \sigma_j^y
+ j\, h_z \sigma_j^z
\right).
\end{equation}
In the strong Stark field regime, the linear potential acts as a kinetic constraint that dominates the spectral properties. Every energy difference of the Stark potential is an integer multiple of $\Omega=2h_z$ (see End Matter).

{\it PXP model---}The PXP model is the paradigmatic example of quantum many-body scarring~\cite{bernien2017probing,turner2018weak}. We consider the chain with open boundary conditions, whose Hamiltonian is given by
\begin{equation}
H = \sum_{j=2}^{L-1} P_{j-1} \sigma_j^x P_{j+1},
\qquad
P_j = \frac{\mathbb{I}-\sigma_j^z}{2}.
\label{eq:pxp_model}
\end{equation}
In contrast to the previous two models, the PXP model achieves spectral commensurability through the Rydberg blockade kinetic constraint. Well-separated excitation patterns span invariant subspaces of the dynamics. Within such a subspace the spectrum forms an integer-valued ladder with unit spacing, yielding $\Omega=1$ (see End Matter).

\begin{figure}[b]
    \centering
    \includegraphics[width=\columnwidth]{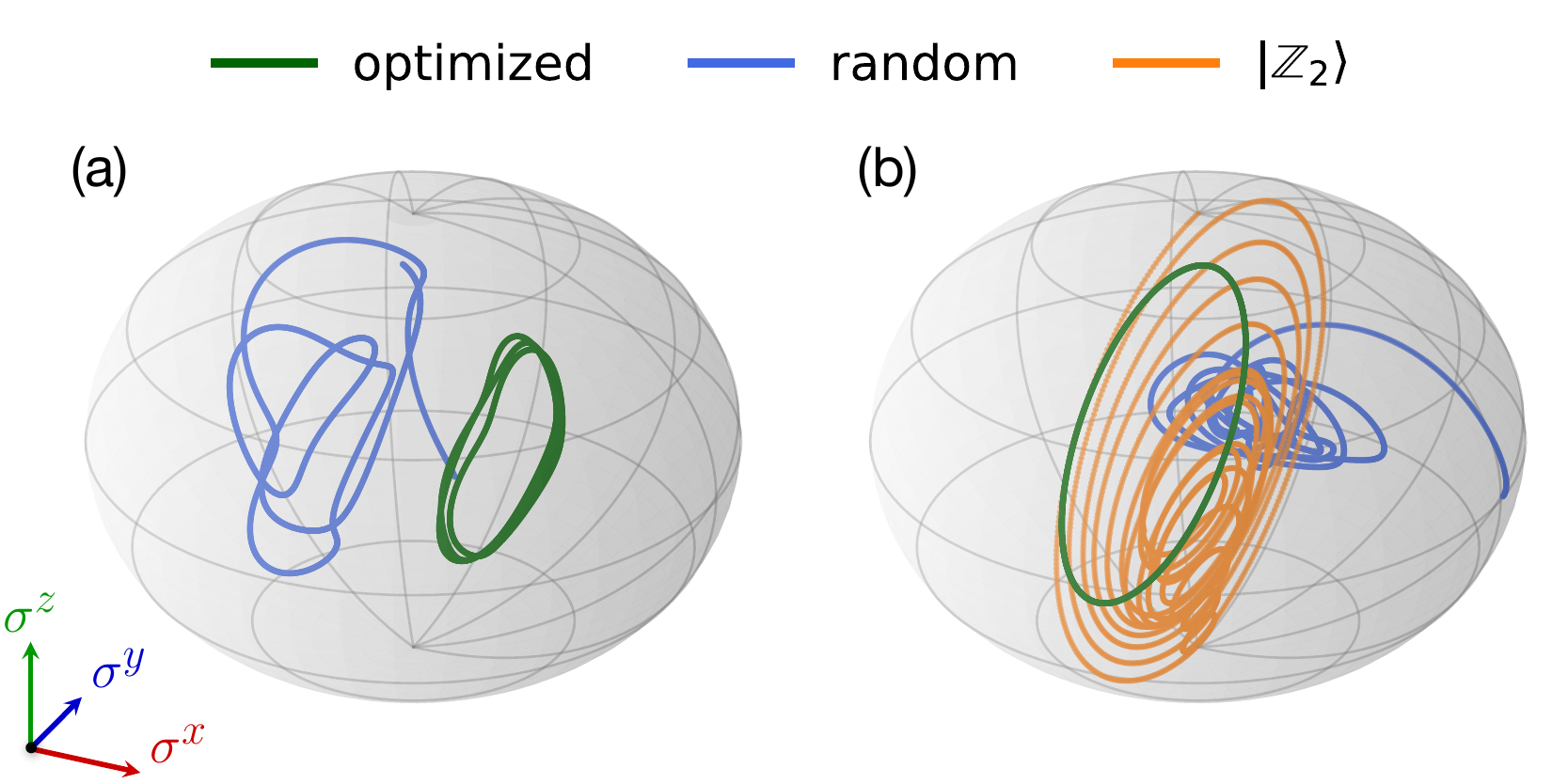}
    \caption{Bloch-sphere trajectories of $(\langle\sigma^x_{11}\rangle,\langle\sigma^y_{11}\rangle,\langle\sigma^z_{11}\rangle)$ under exact time evolution for (a)~the Heisenberg chain and (b)~the PXP model, comparing the optimized Trotter-scar state (green), a Haar-random product state (blue), and the N\'eel state $|\mathbb{Z}_2\rangle$ (orange, PXP only). In both models, the optimized state traces a simple closed orbit, directly reflecting the periodic revivals of the global wave function. By contrast, random states follow irregular trajectories. In the PXP model, the $|\mathbb{Z}_2\rangle$ state follows nested spiraling orbits whose amplitude gradually decays, mirroring the slow degradation of its Loschmidt-echo revivals. Parameters are the same as in Fig.~\ref{fig:trotter_error}.}
    \label{fig:phase_space}
\end{figure}

{\it Results---}We apply the optimization to all three models at $L=12$ with $T_l=10$ and benchmark the optimized states against ensembles of random product-state initial conditions. The most striking result appears in the Loschmidt echo [Figs.~\ref{fig:trotter_error}(a$\sim$c)]. In all three models, the Loschmidt echo of the optimized state displays pronounced periodic revivals throughout the entire evolution, well beyond the optimization window $T_l=10$, with the revival period perfectly matching the analytically predicted stroboscopic time $t_p=2\pi p/\Omega$. In contrast, 100 random product states $|\psi_r\rangle=\bigotimes_j R_j|0\rangle_j$ generated by single-qubit Haar-random gates show markedly different behavior: in the Heisenberg chain, the Loschmidt echo decays rapidly without any revival, while in the Stark chain and PXP model, although revivals are present, their peak values do not return to 1 and gradually diminish over time. Consistent with these Loschmidt echo signatures, the accumulated Trotter error of the optimized state is dramatically suppressed compared with random initial states [Fig.~\ref{fig:trotter_error}(d$\sim$f)]. In the Stark chain, the Trotter error exhibits clear periodic drops at the stroboscopic times during early evolution, reflecting the strong ladder structure of the spectrum. The optimized trajectory closely follows the perturbative prediction of Eq.~\eqref{eq:error_norm_spectral} and remains substantially below the random-state baseline, confirming that the error suppression is a robust feature of Trotter scars rather than an artifact of the training procedure. The time-step dependence of this error is examined in the Supplemental Material~\cite{SM}. 

For the PXP model, we additionally compare the optimized state with the N\'eel state $|\mathbb{Z}_2\rangle = |10\cdots10\rangle$, the prototypical many-body scar initial state known to exhibit persistent revivals under PXP dynamics~\cite{bernien2017probing,turner2018weak}. Although the $|\mathbb{Z}_2\rangle$ state also shows revivals, the revival peaks gradually diminish over time, a decay also reflected in its spiraling Bloch-sphere trajectory [Fig.~\ref{fig:phase_space}(b)]. The accumulated Trotter error [Fig.~\ref{fig:trotter_error}(f)] reveals that the $|\mathbb{Z}_2\rangle$ state is close to the average case, while the optimized state achieves an error roughly six orders of magnitude smaller. This is consistent with our perturbative analysis, where Trotter-error suppression requires both spectral commensurability and favorable error kernel matrix elements, conditions that many-body scar states do not automatically satisfy.

We further implement the Trotter-scar dynamics of all three models on a 17-qubit register of a superconducting processor accessed through the Logical Qubit quantum cloud platform~\cite{logicalqubit}. The device parameters are summarized in the Supplemental Material~\cite{SM}. Figures~\ref{fig:trotter_error}(g$\sim$i) show the measured echo against exact diagonalization for $L=4,6,12$ with the revivals observed at all three sizes. The departure from exact diagonalization grows with the system size, which reflects the accumulation of gate errors with circuit depth (see End Matter for the circuit compilation). 

The signature of persistent revivals is further corroborated by the single-site Bloch-sphere trajectories [Fig.~\ref{fig:phase_space}], where the optimized states trace simple closed orbits in contrast to the irregular trajectories of random initial states. The Supplemental Material shows the corresponding suppression of entanglement entropy~\cite{SM}. All these dynamical features originate from the spectral support of the optimized states~\cite{SM}, which concentrates on equally spaced energy levels, confirming the commensurability condition $\omega_{nm} = k\,\Omega$ required for the $\sin(\omega_{nm} t/2)$ factors in Eq.~\eqref{eq:error_norm_spectral} to vanish simultaneously at the stroboscopic times. Because the stroboscopic factor $\sin(\omega_{nm}t/2)$ is independent of the formula order, this mechanism extends to product formulas of any order, from first-order Lie--Trotter to $(2k)$th-order Suzuki formulas~\cite{SM}.

{\it Discussion.---}We have established a spectral theory of Trotter-error resilience. Initial states supported on commensurate energy ladders exhibit strongly suppressed Trotter errors and persistent Loschmidt revivals, a mechanism that depends only on the spectral structure of $H$ and extends to product formulas of any order. The Loschmidt echo therefore serves as a robust and experimentally accessible indicator of Trotter scars. Guided by this theory, a variational product-state ansatz successfully discovered Trotter scars in three complementary models, the Heisenberg chain with a transverse field, the Stark spin chain, and the PXP model. The existence of Trotter scars shows that Trotter errors can differ sharply from both worst-case and average-case estimates, implying that error budgets in digital quantum simulation should account for the initial state. Symmetries or kinetic constraints that generate spectrally commensurate structures naturally support such error-resilient states, typically within low-dimensional subspaces of the full Hilbert space. Our variational framework is model-agnostic and requires only Trotterized time evolution, making it broadly applicable even when commensurability is only approximate. We have also demonstrated that Trotter scars can be readily prepared on present-day superconducting quantum processors and that the predicted revivals remain visible despite device noise.

{\it Acknowledgments—} This work is supported by Project  12325501, 12047503, and 12247104 of the National Natural Science Foundation of China. P.Z. is partially supported by the Innovation Program for Quantum Science and Technology project 2021ZD0301900. B. Z. acknowledges support from Project 12447153 of the National Natural Science Foundation of China and the China Postdoctoral Science Foundation under Grant Number 2025M773435.

\bibliography{references}

\section*{End Matter}
\setcounter{equation}{0}

\subsection*{Optimization details}

The second term of the loss is weighted by $l_2=10^{-5}$ relative to $l_1=1$. Trivial fixed points of the search are states that remain nearly stationary under Trotterized evolution, such as simultaneous eigenstates of $H$ and $H_{\mathrm{eff}}$. The second term penalizes these states and thereby ensures that the identified Trotter scars correspond to genuinely dynamical trajectories. In practice, $\mathcal{L}$ is minimized using the Adam optimizer with cosine annealing, where gradients with respect to the variational parameters are obtained by backpropagation through the classical simulation of Trotterized dynamics.

\subsection*{Origin of the commensurate ladders}

Here we derive the commensurate ladders of the three models in the main text.

In the Heisenberg chain, $H_{\mathrm{iso}}$ is SU(2) invariant, so the Hilbert space decomposes into multiplets labeled by total spin $S$ and a multiplicity index $\lambda$, with all $2S+1$ states in a multiplet sharing the eigenvalue $E(S,\lambda)$. The transverse field preserves $S$ but lifts this degeneracy through Zeeman-like splitting, yielding the spectrum
\begin{equation}
E_{m_x} = E(S,\lambda) + h_x m_x ,
\qquad
m_x = -S,\ldots,S ,
\label{eq:Heisenberg_ladder}
\end{equation}
forming an exactly equidistant ladder with uniform spacing $\Delta E = h_x$.
The energy differences between any two states within the multiplet are
$\omega_{m_xm_x'} = (m_x - m_x')\, h_x$, i.e., integer multiples of $\Omega=h_x$.
The stroboscopic factor $\sin(\omega_{m_xm_x'} t/2)$ therefore vanishes at stroboscopic times $t_p = 2\pi p / h_x$.

In the Stark chain, consider the Stark potential
$H_0 = h_z \sum_{j=1}^L j \sigma_j^z$.
The energy difference between two computational basis states
$|\mathbf{s}\rangle$ and $|\mathbf{s}'\rangle$ is given by
\begin{equation}
E(\mathbf{s}) - E(\mathbf{s}')
= h_z \sum_{j=1}^L j (s_j - s'_j).
\end{equation}
Since $(s_j - s'_j) \in \{0,\pm 2\}$ and $j$ is a positive integer, every many-body energy difference is an integer multiple of $\Omega=2h_z$, yielding stroboscopic times
$t_p = \pi p / h_z$ .

In the PXP model, each local term $T_j \equiv P_{j-1}\sigma_j^x P_{j+1}$ has eigenvalues restricted to $\{0, \pm 1\}$ and commutes with $T_k$ for $|j-k|>2$. A site $j$ is active when both neighbors satisfy $s_{j\pm 1}=1$. If all active sites are mutually well-separated and isolated by $s_{j\pm 2}=0$, this separation is preserved under the dynamics and the corresponding states form an invariant subspace. Within such a subspace, every active $T_j$ has eigenvalues $\pm 1$ and all inactive terms vanish. The spectrum therefore forms an integer-valued ladder with unit spacing $\Delta E=1$, yielding $\Omega=1$ and stroboscopic times $t_p=2\pi p$.

\subsection*{Hardware implementation}

For the Heisenberg model the bond gate $e^{-i\Delta t\,h_{j,j+1}}$ has Weyl coordinates $c_1=c_2=-c_3=\Delta t/4$ and therefore needs three CZ gates. For the Stark model we synthesize each bond gate with Qiskit's \texttt{TwoQubitBasisDecomposer}~\cite{qiskit}. This approximate Cartan decomposition weighs the truncation error of a shorter circuit against the assumed error of each CZ gate and returns the CZ count with the highest expected fidelity. For the PXP model the three-body term $P_{j-1}\sigma^x_j P_{j+1}$ expands into four mutually commuting Pauli strings, so its exponential factorizes exactly and compiles into four CZ gates.

\clearpage
\onecolumngrid
\begin{center}
\textbf{\large Supplemental Material for ``Trotter Scars: Trotter Error Suppression in Quantum Simulation''}
\end{center}
\setcounter{equation}{0}
\setcounter{figure}{0}
\setcounter{table}{0}
\setcounter{section}{0}
\setcounter{secnumdepth}{3}
\renewcommand{\thefigure}{S\arabic{figure}}
\renewcommand{\theequation}{S\arabic{equation}}
\renewcommand{\thetable}{S\arabic{table}}

\section{Time-step analysis of Trotter-scar states}
\label{sec:dtscan}

This section examines how the Trotter error of the Trotter-scar states
depends on the time step.
The stroboscopic times $t_p=2\pi p/\Omega$ are set by the spectrum of the
Hamiltonian, as explained in the main text.
The time step $\Delta t$ is an independent tunable parameter.
The initial states are the variationally optimized Trotter-scar states of
the main text obtained at $\Delta t=0.01$.
Throughout this section they are loaded as fixed vectors and are not
reoptimized at any step size.

\begin{figure}[h]
    \centering
    \includegraphics[width=0.9\textwidth]{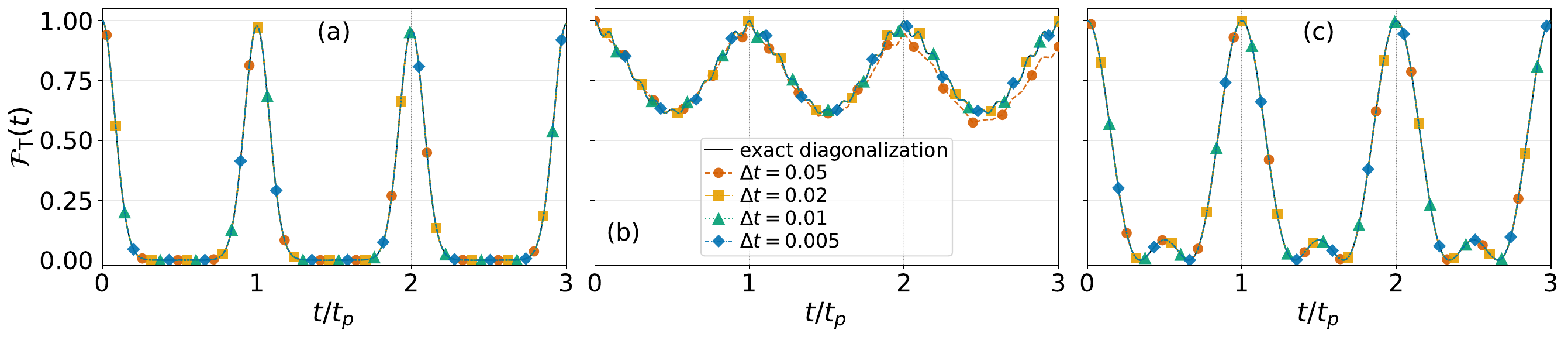}
    \caption{Trotterized Loschmidt echo $\mathcal{F}_{\mathrm{T}}(t)$ of
    the fixed optimized states over three revival periods for (a)~the
    Heisenberg chain, (b)~the Stark chain and (c)~the PXP model.}
    \label{fig:LE_dt}
\end{figure}

For each model the second-order formula $\mathcal{S}_2(\Delta t)$ of
Eq.~(1) in the main text is applied with the step sizes
\begin{equation}
  \Delta t\in\{0.05,\,0.02,\,0.01,\,0.005\}.
  \label{eq:dtscan-values}
\end{equation}
Figure~\ref{fig:LE_dt} follows the resulting Trotterized Loschmidt echo
over three revival periods.
All step sizes reproduce the same revival period $t_1=2\pi/\Omega$.
Its values are $t_1=4\pi$ for the Heisenberg chain, $\pi/4$ for the Stark
chain and $2\pi$ for the PXP model.

\begin{figure}[h!]
    \centering
    \includegraphics[width=0.4\textwidth]{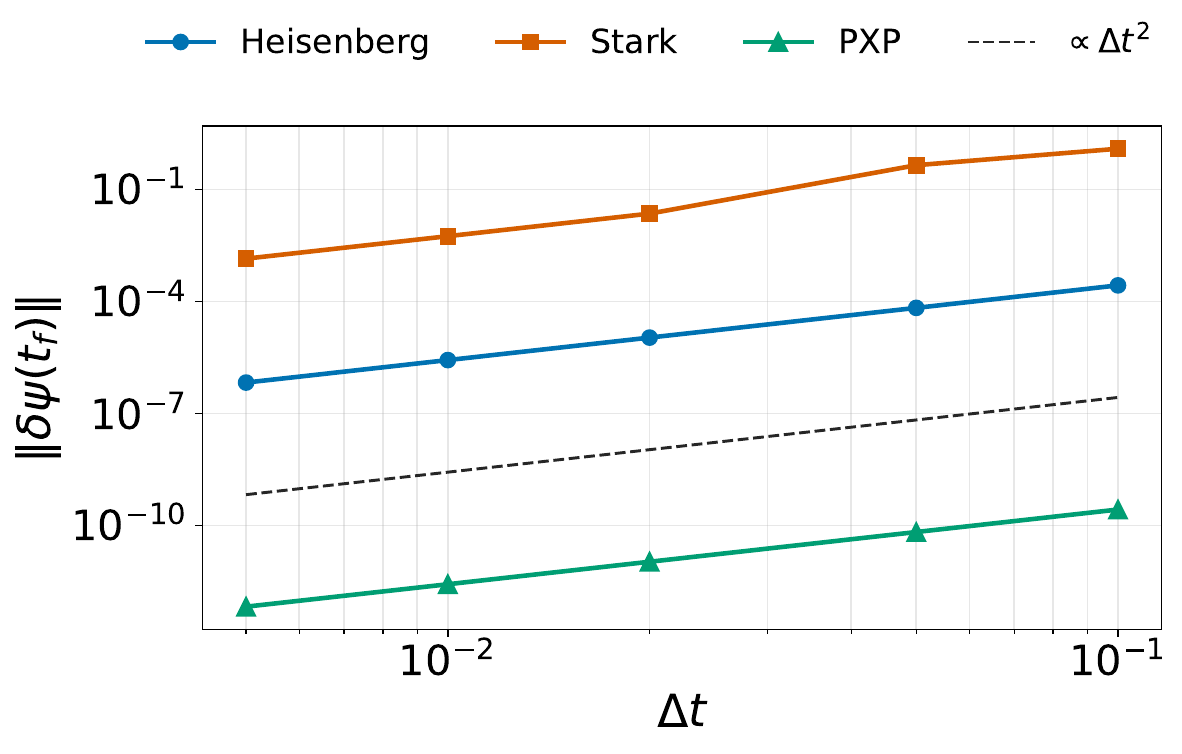}
    \caption{Trotter error at the final time $t_f=20$ versus the time
    step $\Delta t$ for the three optimized Trotter-scar states.
    The optimized initial states are held fixed while the time step $\Delta t$ of the
    second-order formula $\mathcal{S}_2(\Delta t)$ is varied.
    The dashed line indicates the expected $\Delta t^{2}$ scaling.
    The errors approach this scaling asymptotically as $\Delta t$ decreases.}
    \label{fig:dt_conv}
\end{figure}

Figure~\ref{fig:dt_conv} quantifies this step-size dependence.
It shows the accumulated state-vector error $\|\delta\psi(t_f)\|$ as a
function of $\Delta t$.
All models and all step sizes are compared at the same fixed final time
$t_f=20$.
The errors converge asymptotically as $\Delta t^2$, as expected for the
second-order formula.

\section{Entanglement entropy of Trotter-scar states}
\label{app:entropy}

In this section we examine the entanglement dynamics of the Trotter-scar
states.
Figures~2(a) and 2(b) of the main text show that the optimized states
trace simple closed orbits on the single-site Bloch sphere, whereas
Haar-random product states follow irregular trajectories.
Here we establish the many-body counterpart of this single-site
regularity.
The diagnostic is the growth of the half-chain entanglement entropy
under the continuous-time evolution generated by the Hamiltonian
itself.
A state that thermalizes develops extensive entanglement.
A state whose dynamics remains confined to a small subspace does not.

\begin{figure}[h]
    \centering
    \includegraphics[width=\textwidth]{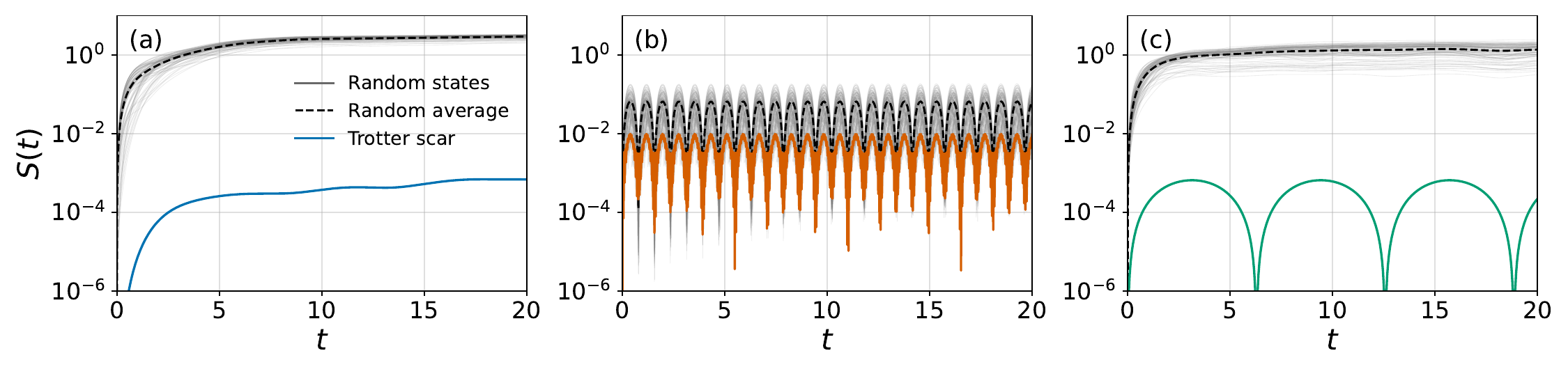}
    \caption{Half-chain entanglement entropy $S(t)$ of the optimized
    Trotter-scar states at $L=12$ under exact diagonalization in
    continuous time for (a)~the Heisenberg chain, (b)~the Stark chain
    and (c)~the PXP model.
    The colored solid line is the optimized Trotter-scar state.
    The thin gray lines are the individual trajectories of $100$
    Haar-random product states evolved under the same Hamiltonian.
    The black dashed line is their average.
    The entropy of the scar states stays at the $10^{-4}$--$10^{-2}$
    level for all simulated times.
    It lies orders of magnitude below the random-state average in the
    Heisenberg chain and the PXP model.
    The gap is much smaller in the Stark chain.
    This reflects the localized character of the model itself.
    Random product states do not thermalize there either.
    Parameters are the same as in Fig.~1 of the main text.}
    \label{fig:ent}
\end{figure}

Figure~\ref{fig:ent} shows the half-chain entanglement entropy $S(t)$
at $L=12$, obtained from exact diagonalization.
For reference, each panel also shows the entropy averaged over $100$
Haar-random product states evolved under the same Hamiltonian.
In the Heisenberg chain and the PXP model the random states rapidly
develop large entanglement with $S\sim O(1)$.
The entropy of the optimized states instead stays at the
$10^{-4}$--$10^{-2}$ level at all simulated times.
In the PXP model the entropy of the optimized state additionally dips
at the revival time $t_1$, where the wave function returns to the
initial product state.
The separation is smaller in the Stark chain.
The chain is localized and random product states fail to thermalize as
well.
Their entropies oscillate in phase because the near-equidistant
Wannier--Stark spectrum drives coherent oscillations for every initial
state.
The entropy of the optimized state nevertheless remains below the
random average.
The dynamics launched from the optimized states therefore remains
effectively confined to the small ladder-supported subspace identified
in the main text.
This is the many-body face of the regular orbits in Fig.~2 of the main
text.

The origin of this confinement is model dependent.
In the PXP model it descends from the kinetically constrained dynamics
of the Rydberg blockade.
The PXP bulk spectrum is nevertheless thermalizing, and random product
states indeed relax toward volume-law entanglement.
This model therefore provides the closest connection to conventional
many-body-scar phenomenology.
In the Heisenberg chain the commensurate ladder instead descends from
the SU(2) multiplet structure.
In the Stark chain it descends from the strong Stark field.
The Trotter-scar mechanism of the main text requires only a
commensurate ladder with favorable error-kernel matrix elements, and it
operates identically in the three settings.

\section{Trotter-scar states at larger system sizes}
\label{app:largeL}

In this section we extend the variational optimization of the main
text to system sizes beyond the reach of exact diagonalization.
The first term of the loss contains the error state
$|\delta\psi(T_l)\rangle$.
Its definition involves the exact propagator $e^{-iHt}$, which is
unavailable at these sizes.
The perturbative analysis provides the way out.
Let $|\psi_{\Delta t}(t)\rangle=[\mathcal{S}_2(\Delta
t)]^{t/\Delta t}|\psi_0\rangle$ denote the Trotterized state at step
$\Delta t$ and $|\delta\psi_{\Delta t}(t)\rangle$ its error state.
Eq.~(5) of the main text shows that the leading error is quadratic in
the step,
\begin{equation}
|\delta\psi_{\Delta t}(t)\rangle=\Delta t^{2}\,|\xi(t)\rangle
+O(\Delta t^{4}),
\qquad
|\xi(t)\rangle=i\int_{0}^{t}d\tau\,
e^{-iH(t-\tau)}K_{2}\,e^{-iH\tau}\,|\psi_{0}\rangle,
\label{eq:xi_state}
\end{equation}
where the vector $|\xi(t)\rangle$ does not depend on $\Delta t$.
Running the same formula at steps $\Delta t$ and $\Delta t/2$ and
subtracting the two trajectories eliminates the exact reference,
\begin{equation}
|\psi_{\Delta t}(t)\rangle-|\psi_{\Delta t/2}(t)\rangle
=|\delta\psi_{\Delta t/2}(t)\rangle-|\delta\psi_{\Delta t}(t)\rangle
=-\frac{3}{4}\,\Delta t^{2}\,|\xi(t)\rangle+O(\Delta t^{4}).
\end{equation}
Taking norms gives the identity
\begin{equation}
\bigl\|\,|\psi_{\Delta t}(t)\rangle-|\psi_{\Delta t/2}(t)\rangle\,\bigr\|
=\frac{3}{4}\,\bigl\|\delta\psi_{\Delta t}(t)\bigr\|+O(\Delta t^{4}).
\label{eq:proxy}
\end{equation}
Then we can replace the first term of the loss function in the main
text by the norm of the trajectory difference,
\begin{equation}
\mathcal{L}_{A}(\boldsymbol{\theta},\boldsymbol{\phi})
=l_{1}\,\bigl\|\,|\psi_{\Delta t}(T_l)\rangle
-|\psi_{\Delta t/2}(T_l)\rangle\,\bigr\|
+\frac{l_{2}}{T_l}\int_{0}^{T_l}\mathcal{F}_{\mathrm{T}}(t)\,dt.
\label{eq:lossA}
\end{equation}
The corresponding factor $4/3$ is absorbed into a redefinition of $l_1$.
The echo term involves Trotterized states alone and needs no
modification.
The quartic corrections in Eq.~\eqref{eq:proxy} are controlled by the
step-size analysis of Fig.~\ref{fig:dt_conv}.
The errors of all three optimized states follow the expected
$\Delta t^{2}$ law asymptotically.

This loss function allows us to evolve both trajectories as full
state vectors.
For larger sizes they can be evolved as matrix product states with
the time-evolving block decimation algorithm at bond dimension $\chi$.
Gradients with respect to the $2L$ variational angles are obtained by
backpropagation through the tensor contractions.
The optimizer and the annealing schedule follow the End Matter of
the main text.

The remaining cost of the optimization lies in the number of gradient
iterations.
To further reduce this cost we employ explicitly constructed initial
states guided by the structure discovered variationally at small sizes.
The variational optima at small sizes reveal structure that
extrapolates in $L$.
Turning this structure into an explicit construction gives each model a
deterministic initial state.
A short refinement is then optional rather than necessary.
The underlying structure differs between the three models as follows.

\begin{itemize}
\item \emph{Heisenberg chain.}
The optimized single-site angles form a smooth profile in the scaled
coordinate $j/L$.
We therefore interpolate the profile of a smaller optimized chain onto
the larger lattice.
The angles are first folded to a single chart of the Bloch sphere to
remove the gauge redundancy of the parameterization.

\item \emph{Stark chain.}
Every site of the optimized state points close to one of the two poles
of the Bloch sphere.
The tilt of site $j$ away from its pole decreases along the chain,
since the growing Stark field pins the spins ever more strongly.
The tilt follows the power law
$\delta_{j}=\delta_{0}\,j^{-\gamma}$ with $\delta_{0}=0.404$ and
$\gamma=1.145$.
Both parameters are fitted once from the small-size optima.
For system sizes that are a multiple of twelve we repeat the pole and
azimuth pattern of the $L=12$ optimum,
\begin{equation}
\theta_{j}=
\begin{cases}
\delta_{j}\,, & s^{(12)}_{i}=+\,,\\
\pi-\delta_{j}\,, & s^{(12)}_{i}=-\,,
\end{cases}
\qquad
\phi_{j}=\phi^{(12)}_{i},
\qquad
i=(j-1)\bmod 12,
\label{eq:stark_construction}
\end{equation}
where $s^{(12)}_{i}$ and $\phi^{(12)}_{i}$ are the pole pattern and the
azimuths read off from the $L=12$ optimum.

\item \emph{PXP model.}
The optimized states consist of isolated three-site cages on a
$\ket{0}$ background.
A cage is a group of three adjacent sites with nontrivial Bloch angles.
The Rydberg blockade prevents the cages from interacting across the
intervening $\ket{0}$ stretches.
We harvest the cage angles from the $L=24$ optimum and repeat the cages
along the larger chain.
The construction involves no optimization at any size.
\end{itemize}

\begin{figure}[h]
    \centering
    \includegraphics[width=\textwidth]{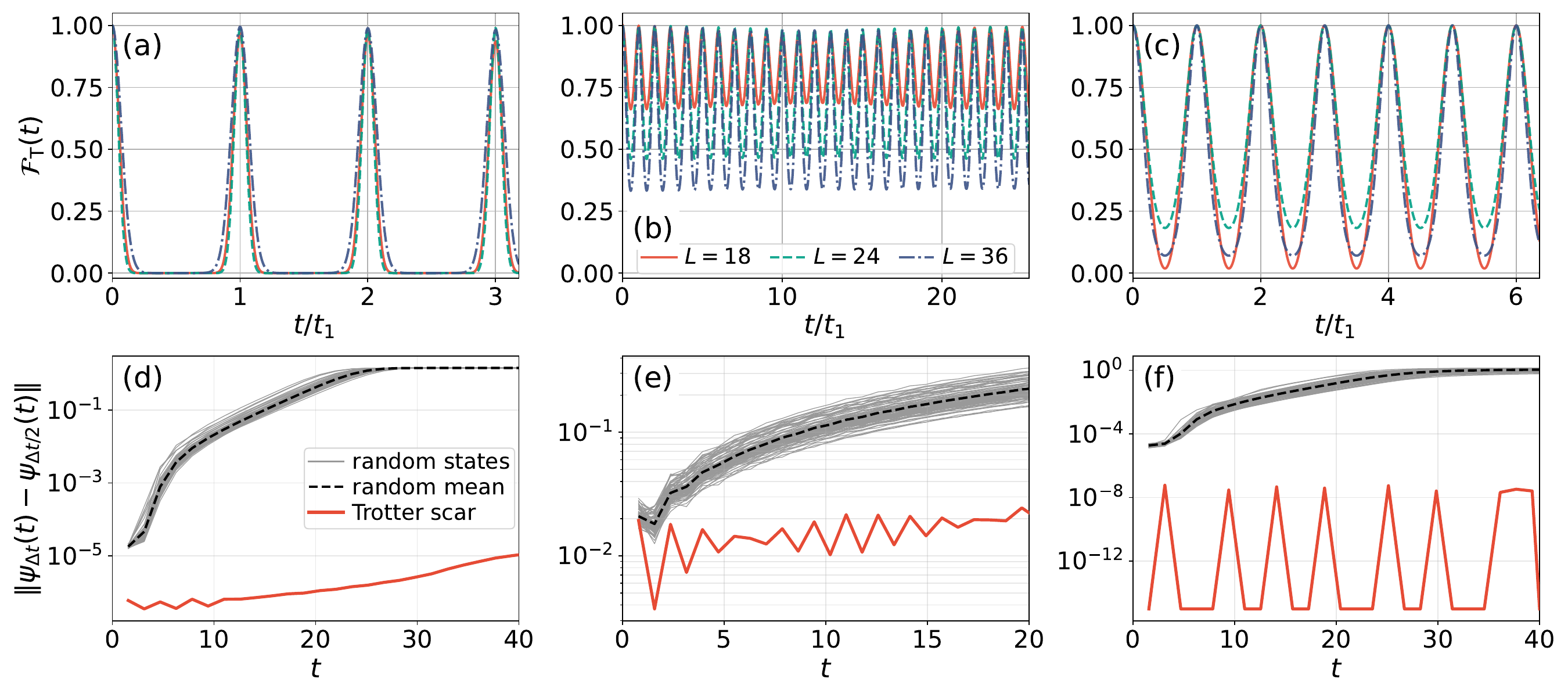}
    \caption{Trotter-scar states for (a,d)~the Heisenberg chain,
    (b,e)~the Stark chain and (c,f)~the PXP model.
    Panels (a)--(c) show the Trotterized Loschmidt echo
    $\mathcal{F}_{\mathrm{T}}(t)$ at $\Delta t=0.01$ for system sizes
    $L\in\{18,24,36\}$.
    The time axis is in units of the revival period $t_1$.
    The $L=18$ and $L=24$ states are variationally optimized with the
    loss of Eq.~\eqref{eq:lossA}.
    The $L=36$ states are obtained from the deterministic
    constructions described in the text.
    Only the Heisenberg state at $L=36$ includes a short refinement
    after interpolation.
    All other constructed states involve no optimization.
    The peak positions remain locked to the stroboscopic times at every
    size.
    Panels (d)--(f) show the trajectory difference
    $\|\,|\psi_{\Delta t}(t)\rangle-|\psi_{\Delta t/2}(t)\rangle\,\|$
    of the $L=36$ states.
    By Eq.~\eqref{eq:proxy} this measures the Trotter error at
    leading order up to the factor $3/4$.
    The red line is the Trotter-scar state.
    The thin gray lines are $100$ Haar-random product states evolved
    with the same Trotter circuits.
    The black dashed line is their average.}
    \label{fig:LE_sizes}
\end{figure}

Figure~\ref{fig:LE_sizes} shows the results for the three models.
The upper panels show the Trotterized Loschmidt echo at
$\Delta t=0.01$ for $L\in\{18,24,36\}$.
The $L=18$ and $L=24$ states are variationally optimized with the loss
of Eq.~\eqref{eq:lossA}.
The $L=18$ states and the $L=24$ PXP state are optimized with full
state vectors on a GPU.
The Heisenberg and Stark states at $L=24$ are optimized as matrix
product states at $\chi=8$.
The $L=36$ states follow the deterministic constructions above.
Only the Heisenberg state at $L=36$ receives a short refinement at
$\chi=32$ with fixed hyperparameters.
Every other constructed state enters the figure without refinement.
All revival trains persist over the full windows.
The peak positions stay locked to the stroboscopic times
$t_p=2\pi p/\Omega$ at every size.
The lower panels compare the $L=36$ states with $100$ Haar-random
product states evolved by the same Trotter circuits.
The plotted quantity is the trajectory difference
$\|\,|\psi_{\Delta t}(t)\rangle-|\psi_{\Delta t/2}(t)\rangle\,\|$.
This is the first term of the loss in Eq.~\eqref{eq:lossA}.
By Eq.~\eqref{eq:proxy} it measures the Trotter error
$\|\delta\psi_{\Delta t}(t)\|$ at leading order up to the factor $3/4$.
The random-state error accumulates with time.
The error of the Trotter-scar states shows no comparable growth.

\newpage
\section{Perturbative analysis for higher-order product formulas}
\label{app:higher_order}

This section provides the detailed derivations of the perturbative analysis for product formulas beyond the second-order case presented in the main text. We first review the second-order case and construct the fourth-order Suzuki formula explicitly. We prove by induction that the error suppression mechanism is universal across all even-order Suzuki formulas. We then extend the analysis to the first-order Lie--Trotter decomposition in the next section and show that the same stroboscopic factor governs the error suppression.

\subsection{Second-order decomposition}
\label{app:second_order}

\subsubsection{BCH expansion and symmetric splitting}

The second-order symmetric Trotter decomposition
\begin{equation}
\mathcal{S}_2(\Delta t) = e^{-iH_o\Delta t/2}\,e^{-iH_e\Delta t}\,e^{-iH_o\Delta t/2}
\end{equation}
possesses a symmetric structure ($H_o/2,\,H_e,\,H_o/2$).
We now prove that this symmetry constrains the BCH expansion to contain only odd powers of $\Delta t$.

Reversing $\Delta t\to -\Delta t$ in each exponential factor gives
\begin{equation}
\mathcal{S}_2(-\Delta t) = e^{iH_o\Delta t/2}\,e^{iH_e\Delta t}\,e^{iH_o\Delta t/2}.
\end{equation}
Meanwhile, inverting $\mathcal{S}_2(\Delta t)$ reverses the order and conjugates each factor:
\begin{equation}
\mathcal{S}_2(\Delta t)^{-1} = \bigl(e^{-iH_o\Delta t/2}\,e^{-iH_e\Delta t}\,e^{-iH_o\Delta t/2}\bigr)^{-1} = e^{iH_o\Delta t/2}\,e^{iH_e\Delta t}\,e^{iH_o\Delta t/2}.
\end{equation}
These are identical, establishing the key identity $\mathcal{S}_2(-\Delta t) = \mathcal{S}_2(\Delta t)^{-1}$.

Writing the generator as a formal power series $\ln \mathcal{S}_2(\Delta t) = \sum_{n=1}^{\infty} a_n\,\Delta t^n$,
the identity $\ln \mathcal{S}_2(-\Delta t) = -\ln \mathcal{S}_2(\Delta t)$ becomes
$\sum_n (-1)^n a_n\,\Delta t^n = -\sum_n a_n\,\Delta t^n$.
Comparing order by order: $a_n\bigl((-1)^n+1\bigr)=0$.
For even $n$, $(-1)^n+1=2\neq 0$, forcing $a_n=0$; for odd $n$, the condition is automatically satisfied.
Therefore, the BCH expansion contains only odd powers of $\Delta t$:
\begin{equation}
\ln \mathcal{S}_2(\Delta t) = -i\bigl(H\,\Delta t + W_3\,\Delta t^3 + W_5\,\Delta t^5 + \cdots\bigr),
\label{eq:bch_odd}
\end{equation}
To compute $W_3$ explicitly, we apply the BCH formula to $\mathcal{S}_2=e^{A}e^{B}e^{A}$ with $A=-iH_o\Delta t/2$ and $B=-iH_e\Delta t$ in two steps: first combining $e^A e^B = e^D$, then $e^D e^A$.
At $O(\Delta t)$, the sum recovers $-iH\,\Delta t$. The $O(\Delta t^2)$ terms from $D$ and $\frac{1}{2}[D,A]$ cancel exactly, consistent with the symmetry argument above. Collecting all $O(\Delta t^3)$ contributions, including direct terms from the first BCH step, cross-terms between $O(\Delta t^2)$ and $O(\Delta t)$ pieces, and the $\frac{1}{12}$ nested commutators, yields the leading error operator
\begin{equation}
W_3 \equiv K_2 = \tfrac{1}{24}\bigl([H_o,[H_o,H_e]] + 2[H_e,[H_o,H_e]]\bigr).
\end{equation}

\subsubsection{Effective Hamiltonian}

The single-step propagator $\mathcal{S}_2(\Delta t)$ is unitary, so it can always be written as the exponential of some Hermitian generator:
$\mathcal{S}_2(\Delta t) \equiv e^{-iH_{\mathrm{eff}}\,\Delta t}$.
This defines the effective Hamiltonian $H_{\mathrm{eff}}$.
Taking the logarithm of both sides and comparing with Eq.~\eqref{eq:bch_odd}:
\begin{equation}
-iH_{\mathrm{eff}}\,\Delta t = -i\bigl(H\,\Delta t + W_3\,\Delta t^3 + W_5\,\Delta t^5 + \cdots\bigr).
\end{equation}
Dividing both sides by $-i\Delta t$ converts the odd powers $(\Delta t,\,\Delta t^3,\,\Delta t^5,\ldots)$ into even powers $(1,\,\Delta t^2,\,\Delta t^4,\ldots)$:
\begin{equation}
H_{\mathrm{eff}} = H + \Delta t^2 K_2 + O(\Delta t^4).
\label{eq:heff_second}
\end{equation}
An important consequence is that multi-step Trotter evolution is \emph{exact} at the level of $H_{\mathrm{eff}}$:
since $(e^X)^N=e^{NX}$ holds for any matrix $X$ (which trivially commutes with itself), we have
$[\mathcal{S}_2(\Delta t)]^N = e^{-iH_{\mathrm{eff}}\cdot N\Delta t} = e^{-iH_{\mathrm{eff}}\,t}$
with $t=N\Delta t$.
The only approximation lies in the BCH truncation of $H_{\mathrm{eff}}$ itself.

\subsubsection{Interaction-picture perturbation theory}

We now derive the error state systematically.
Writing $H_{\mathrm{eff}}=H+V$ with $V=\Delta t^2 K_2$, and defining the interaction-picture propagator
\begin{equation}
U_I(t) \equiv e^{iHt}\,e^{-i(H+V)t},
\end{equation}
one can verify by direct differentiation that $U_I(t)$ satisfies the equation of motion
\begin{equation}
i\frac{d}{dt}U_I(t) = V_I(t)\,U_I(t), \qquad V_I(t)\equiv e^{iHt}\,V\,e^{-iHt},
\label{eq:UI_eom}
\end{equation}
with initial condition $U_I(0)=\mathbf{1}$.
The formal solution is the time-ordered Dyson series.
To first order in $V$:
\begin{equation}
U_I(t) \approx \mathbf{1} - i\int_0^t d\tau\, V_I(\tau).
\label{eq:dyson_first}
\end{equation}
Returning to the Schr\"odinger picture by left-multiplying with $e^{-iHt}$:
\begin{equation}
e^{-i(H+V)t} = e^{-iHt}\,U_I(t) \approx e^{-iHt} - i\int_0^t d\tau\, e^{-iH(t-\tau)}\,V\,e^{-iH\tau}.
\end{equation}
The error state $|\delta\psi(t)\rangle \equiv (e^{-iHt}-e^{-iH_{\mathrm{eff}}t})|\psi_0\rangle = (e^{-iHt}-e^{-i(H+V)t})|\psi_0\rangle$ therefore becomes
\begin{equation}
|\delta\psi(t)\rangle \approx i\,\Delta t^2 \int_0^t d\tau\, e^{-iH(t-\tau)}\,K_2\,e^{-iH\tau}\,|\psi_0\rangle,
\label{eq:error_state_app}
\end{equation}
which is Eq.~(5) of the main text.

\subsubsection{Spectral expansion and the stroboscopic factor}

We now expand Eq.~\eqref{eq:error_state_app} in the eigenbasis of $H$.
Let $H|n\rangle=E_n|n\rangle$ and $|\psi_0\rangle=\sum_m c_m|m\rangle$.
Inserting the completeness relation $\sum_n|n\rangle\langle n|$ between $K$ and $e^{-iH\tau}$:
\begin{equation}
|\delta\psi(t)\rangle \approx i\,\Delta t^2 \sum_{n,m} c_m\,(K_2)_{nm}\,|n\rangle\,e^{-iE_n t}\int_0^t d\tau\, e^{i(E_n-E_m)\tau},
\end{equation}
where $K_{nm}\equiv\langle n|K|m\rangle$. The time integral evaluates to
\begin{equation}
\int_0^t d\tau\,e^{i\omega_{nm}\tau} = \frac{e^{i\omega_{nm}t}-1}{i\omega_{nm}} = e^{i\omega_{nm}t/2}\,\frac{2\sin(\omega_{nm}t/2)}{\omega_{nm}},
\label{eq:time_integral}
\end{equation}
where $\omega_{nm}\equiv E_n - E_m$. Substituting back and taking the squared norm (the cross terms between different $|n\rangle$ vanish by orthogonality):
\begin{equation}
\|\delta\psi(t)\|^2 \approx \Delta t^4 \sum_{n}\left|\sum_{m} c_m\,(K_2)_{nm}\, e^{\frac{i\omega_{nm}t}{2}}\,\frac{2\sin\frac{\omega_{nm}t}{2}}{\omega_{nm}}\right|^2,
\end{equation}
which is Eq.~(6) of the main text.

The stroboscopic factor $\sin(\omega_{nm}t/2)$ controls whether each spectral contribution accumulates or is periodically suppressed, while $1/\omega_{nm}$ provides an amplitude weight favoring small energy differences.
For a generic initial state, the error grows as $\|\delta\psi(t)\|^2\propto \Delta t^4 t^2$ at intermediate times.
However, when the initial state is supported on a spectral ladder with commensurate gaps $\omega_{nm}=k\Omega$ ($k\in\mathbb{Z}$), all stroboscopic factors vanish simultaneously at the times $t_p=2\pi p/\Omega$:
\begin{equation}
\sin(k\Omega\, t_p/2) = \sin(kp\pi) = 0 \quad\text{for } k,p\in\mathbb{Z},
\end{equation}
producing error suppression at the stroboscopic times.

\subsection{Fourth-order Suzuki formula}
\label{app:fourth_order}

\subsubsection{Recursive construction}

Suzuki's recursive formula constructs a $(2k{+}2)$-order formula from a $2k$-order one:
\begin{equation}
\mathcal{S}_{2k+2}(\Delta t)=\bigl[\mathcal{S}_{2k}(p_k\Delta t)\bigr]^2 \mathcal{S}_{2k}(s_k\Delta t)\bigl[\mathcal{S}_{2k}(p_k\Delta t)\bigr]^2,
\label{eq:suzuki_recursive}
\end{equation}
with $s_k\equiv 1-4p_k$ and
\begin{equation}
p_k=\frac{1}{4-4^{1/(2k+1)}}.
\end{equation}
The parameters satisfy two conditions: $4p_k+s_k=1$ (consistency) and $4p_k^{2k+1}+s_k^{2k+1}=0$ (leading-error cancellation).

For $k=1$ (constructing $\mathcal{S}_4$ from $\mathcal{S}_2$), we write $p\equiv p_1$ and $s\equiv s_1$ for brevity, giving
$p=1/(4-4^{1/3})\approx 0.4145$ and $s=1-4p\approx -0.6580$.
The resulting gate sequence consists of five $\mathcal{S}_2$ calls:
\begin{equation}
\mathcal{S}_4(\Delta t) = \mathcal{S}_2(p\Delta t)^2\;\mathcal{S}_2(s\Delta t)\;\mathcal{S}_2(p\Delta t)^2,
\label{eq:s4_explicit}
\end{equation}
which expands to an 11-layer symmetric sequence of alternating $e^{-iH_o(\cdot)}$ and $e^{-iH_e(\cdot)}$ gates.

\subsubsection{Cancellation of the $\Delta t^2$ error}

We now show in detail that the $\Delta t^3$ term in $\ln \mathcal{S}_4(\Delta t)$ vanishes, implying the absence of the $\Delta t^2$ correction in $H_{\mathrm{eff}}^{(4)}$.
Since $\mathcal{S}_2(\tau)$ commutes with itself, $\mathcal{S}_2(\tau)^2 = e^{2\ln \mathcal{S}_2(\tau)}$.
Using Eq.~\eqref{eq:bch_odd}, we can write $\mathcal{S}_4$ as a product of three exponential factors:
\begin{equation}
\mathcal{S}_4(\Delta t) = e^{2A}\,e^{B}\,e^{2A},
\end{equation}
where
\begin{align}
2A &= -i\bigl(2pH\,\Delta t + 2p^3 W_3\,\Delta t^3 + O(\Delta t^5)\bigr), \label{eq:factorA}\\
B  &= -i\bigl(s H\,\Delta t + s^3 W_3\,\Delta t^3 + O(\Delta t^5)\bigr). \label{eq:factorB}
\end{align}
Note that each factor's generator at leading order ($O(\Delta t)$) is proportional to the \emph{full} Hamiltonian $H=H_o+H_e$; the non-trivial commutator structure between $H_o$ and $H_e$ is entirely contained within the higher-order terms ($W_3, W_5, \ldots$) internal to each $\mathcal{S}_2$.

We evaluate $\ln(e^{2A}\,e^{B}\,e^{2A})$ by applying the BCH formula in two successive steps.
Recall that for any two operators $X$ and $Y$, the BCH formula reads
\begin{equation}
\ln(e^X e^Y) = X + Y + \tfrac{1}{2}[X,Y] + \tfrac{1}{12}\bigl([X,[X,Y]]+[Y,[Y,X]]\bigr) + \cdots,
\label{eq:bch_formula}
\end{equation}
where the higher-order terms involve increasingly nested commutators of $X$ and $Y$.

Write the generators in Eqs.~\eqref{eq:factorA}--\eqref{eq:factorB} as
\begin{equation}
2A = X_1\,\Delta t + X_3\,\Delta t^3 + O(\Delta t^5), \qquad B = Y_1\,\Delta t + Y_3\,\Delta t^3 + O(\Delta t^5),
\end{equation}
where $X_1=-2ipH$, $X_3=-2ip^3 W_3$, $Y_1=-is H$, $Y_3=-is^3 W_3$.
The crucial observation is that $X_1$ and $Y_1$ are both proportional to the \emph{same} operator $H$.

Now combine the first two factors: $e^{2A}\,e^{B} = e^{C}$, with $C=\ln(e^{2A}\,e^{B})$.
Applying Eq.~\eqref{eq:bch_formula} and collecting by powers of $\Delta t$:

\emph{$O(\Delta t)$:}
\begin{equation}
C^{(1)} =  X_1\,\Delta t + Y_1\,\Delta t = -i(2p+s)H\,\Delta t.
\end{equation}

\emph{$O(\Delta t^2)$:} The BCH commutator $\frac{1}{2}[2A,B]$ at leading order gives
\begin{equation}
\tfrac{1}{2}[X_1\,\Delta t,\,Y_1\,\Delta t] = \tfrac{1}{2}(-i)^2\cdot 2p\cdot s\,[H,H]\,\Delta t^2 = 0.
\end{equation}

\emph{$O(\Delta t^3)$:} Three contributions arise at this order:

(a)~Direct terms from each factor: $X_3\,\Delta t^3 + Y_3\,\Delta t^3 = -i(2p^3 + s^3)W_3\,\Delta t^3$.

(b)~The third-order BCH nested commutators from the $\frac{1}{12}$ term:
\begin{equation}
\tfrac{1}{12}\bigl([X_1\,\Delta t,[X_1\,\Delta t,Y_1\,\Delta t]]+[Y_1\,\Delta t,[Y_1\,\Delta t,X_1\,\Delta t]]\bigr) \propto [H,[H,H]] = 0.
\end{equation}

Therefore
\begin{equation}
C = -i(2p+s)H\,\Delta t - i(2p^3+s^3)W_3\,\Delta t^3 + O(\Delta t^4).
\label{eq:C_intermediate}
\end{equation}

Next, combine with the remaining factor: $e^C\,e^{2A} = e^D$, with $D=\ln(e^C\,e^{2A})$.
The generator $C$ from Eq.~\eqref{eq:C_intermediate} has the same structure as $2A$: both have $O(\Delta t)$ terms proportional to $H$. Applying BCH again:

\emph{$O(\Delta t)$:}
\begin{equation}
D^{(1)} = C^{(1)} + X_1\,\Delta t = -i(2p + s + 2p)H\,\Delta t = -i(4p+s)H\,\Delta t = -iH\,\Delta t,
\end{equation}
using the consistency condition $4p+s=1$.

\emph{$O(\Delta t^2)$:}
\begin{equation}
\tfrac{1}{2}[C^{(1)},\,X_1\,\Delta t] = \tfrac{1}{2}(-i)^2\cdot(2p+s)\cdot 2p\,[H,H]\,\Delta t^2 = 0.
\end{equation}

\emph{$O(\Delta t^3)$:} Again two types of contributions:

(a)~The $O(\Delta t^3)$ pieces from $C$ and $2A$:
\begin{equation}
-i(2p^3+s^3)W_3\,\Delta t^3 + X_3\,\Delta t^3 = -i(2p^3+s^3+2p^3)W_3\,\Delta t^3 = -i(4p^3+s^3)W_3\,\Delta t^3.
\end{equation}
Since $s=-4^{1/3}p$, we have $s^3=-4p^3$, giving
\begin{equation}
4p^3+s^3 = 4p^3 - 4p^3 = 0.
\end{equation}

(b)~Third-order BCH nested commutators from the $\frac{1}{12}$ term:
\begin{equation}
\tfrac{1}{12}\bigl([C^{(1)},[C^{(1)},X_1\,\Delta t]] + [X_1\,\Delta t,[X_1\,\Delta t,C^{(1)}]]\bigr) \propto [H,[H,H]] = 0.
\end{equation}
As before, since $C^{(1)}\propto H$ and $X_1\propto H$, all nested commutators between the leading-order terms of different factors are built entirely from $H$, and $[H,H]=0$ makes the innermost commutator vanish, annihilating the entire expression.

There are no other contributions at $O(\Delta t^3)$: cross-terms mixing $O(\Delta t)$ with $O(\Delta t^2)$ would require even-order terms in some factor, which are absent by the symmetric splitting.
Combining (a) and (b), the $O(\Delta t^3)$ term of $D$ vanishes. It remains to verify that the $O(\Delta t^4)$ term also vanishes.

\emph{$O(\Delta t^4)$:} Two contributions arise.

(a)~The $O(\Delta t^4)$ term inherited from $C$. From the first BCH step, $\frac{1}{2}[2A,B]$ at $O(\Delta t^4)$ gives
\begin{equation}
C^{(4)} = \tfrac{1}{2}\bigl[X_1\,\Delta t,\,Y_3\,\Delta t^3\bigr] + \tfrac{1}{2}\bigl[X_3\,\Delta t^3,\,Y_1\,\Delta t\bigr]
= -p\,s(s^2-p^2)\,[H,W_3]\,\Delta t^4.
\end{equation}

(b)~The cross-commutator $\frac{1}{2}[C,2A]$ at $O(\Delta t^4)$:
\begin{equation}
\tfrac{1}{2}\bigl[C^{(1)},\,X_3\,\Delta t^3\bigr] + \tfrac{1}{2}\bigl[C^{(3)}\,\Delta t^3,\,X_1\,\Delta t\bigr]
= p\,s(s^2-p^2)\,[H,W_3]\,\Delta t^4,
\end{equation}
where $C^{(3)}=-i(2p^3+s^3)W_3\,\Delta t^3$ is the $O(\Delta t^3)$ part of $C$.

The two contributions cancel exactly:
\begin{equation}
D^{(4)} = C^{(4)} + \tfrac{1}{2}[C,2A]^{(4)} = -p\,s(s^2-p^2)\,[H,W_3]\,\Delta t^4 + p\,s(s^2-p^2)\,[H,W_3]\,\Delta t^4 = 0.
\end{equation}
This cancellation is guaranteed by the same time-reversal argument used for $\mathcal{S}_2$: since $\mathcal{S}_4$ is built from symmetric factors, $\mathcal{S}_4(-\Delta t)=\mathcal{S}_4(\Delta t)^{-1}$, so $\ln\mathcal{S}_4$ contains only odd powers of $\Delta t$. We therefore conclude
\begin{equation}
D = \ln \mathcal{S}_4(\Delta t) = -i\bigl(H\,\Delta t + K_4\,\Delta t^5 + O(\Delta t^7)\bigr),
\end{equation}
where $K_4$ is a Hermitian operator (since $D$ is anti-Hermitian and only odd powers survive). The $O(\Delta t^3)$ term vanishes by the Suzuki condition $4p^3+s^3=0$, and the $O(\Delta t^4)$ term vanishes by time-reversal symmetry; the leading error is therefore pushed to $O(\Delta t^5)$.

\subsubsection{Effective Hamiltonian and spectral representation}

The fourth-order effective Hamiltonian is therefore
\begin{equation}
H_{\mathrm{eff}}^{(4)} = H + \Delta t^4 K_4 + O(\Delta t^6),
\label{eq:heff4}
\end{equation}
which follows from dividing the above by $-i\Delta t$. Collecting all $O(\Delta t^5)$ contributions from the two-step BCH, namely the internal $W_5$ terms from each $\mathcal{S}_2$ factor and the cross-commutators between factors, and dividing by $-i\Delta t$, one obtains
\begin{equation}
K_4 = (4p^5+s^5)\,W_5 - \frac{p\,s(2p+s)(p^2-s^2)}{3}\,[H,[H,W_3]].
\label{eq:K4}
\end{equation}
The first term is the direct fifth-order error from each symmetric factor, summed with coefficient $(4p^5+s^5)\neq 0$ (unlike the third order, this combination does not vanish). The second term arises from BCH cross-commutators between the $W_3$ and $H$ contributions of different factors. Fifth-order nested commutators among the leading $H\,\Delta t$ terms, of the form $[H,[H,[H,[H,H]]]]$, vanish identically by $[H,H]=0$.

The interaction-picture perturbation theory with $V=\Delta t^4 K_4$ proceeds identically to the second-order case derived above, with the systematic substitution $\Delta t^2 K_2\to\Delta t^4 K_4$.
The error state is
\begin{equation}
|\delta\psi(t)\rangle \approx i\,\Delta t^4\int_0^t d\tau\,e^{-iH(t-\tau)}\,K_4\,e^{-iH\tau}\,|\psi_0\rangle,
\end{equation}
and expanding in the eigenbasis as before, the squared error norm becomes
\begin{equation}
\|\delta\psi(t)\|^2 \approx \Delta t^{8}\sum_n\left|\sum_m c_m(K_4)_{nm}\,e^{\frac{i\omega_{nm}t}{2}}\frac{2\sin\frac{\omega_{nm}t}{2}}{\omega_{nm}}\right|^2.
\label{eq:error_s4}
\end{equation}
Comparing with Eq.~(6) of the main text, the stroboscopic factor $\sin(\omega_{nm}t/2)$ is \emph{identical}; only the overall scaling ($\Delta t^8$ vs.\ $\Delta t^4$) and the error-kernel matrix elements ($(K_4)_{nm}$ vs.\ $K_{nm}$) differ. The error suppression mechanism therefore carries over unchanged from second to fourth order. We now show that this conclusion holds for Suzuki formulas of arbitrary even order.

\subsection{Generalization to arbitrary even order and universality}
\label{app:universality}

We prove by mathematical induction that for every $k\geq 1$, the $(2k)$th-order Suzuki formula $\mathcal{S}_{2k}$ has an effective Hamiltonian of the form
\begin{equation}
H_{\mathrm{eff}}^{(2k)} = H + \Delta t^{2k}K_{2k}+O(\Delta t^{2k+2}),
\label{eq:heff_general}
\end{equation}
where $K_{2k}$ is a Hermitian operator composed of $(2k{+}1)$-th order nested commutators of $H_o$ and $H_e$.

\emph{Base case} ($k=1$): We have established in the second-order analysis that $H_{\mathrm{eff}}^{(2)} = H + \Delta t^2 K_2 + O(\Delta t^4)$.

\emph{Inductive hypothesis:} Assume that $\mathcal{S}_{2k}$ is a $2k$-order formula whose generator has the structure
\begin{equation}
\ln \mathcal{S}_{2k}(\tau) = -i\bigl(H\tau + C_{2k+1}\,\tau^{2k+1}+O(\tau^{2k+3})\bigr),
\label{eq:inductive_hypothesis}
\end{equation}
where only odd powers of $\tau$ appear (guaranteed by the symmetric splitting of $\mathcal{S}_{2k}$, which holds at every level of the recursion), and $C_{2k+1}$ is the leading error operator.

\emph{Inductive step} ($k\to k{+}1$): The Suzuki recursion~\eqref{eq:suzuki_recursive} constructs $\mathcal{S}_{2k+2}$ from five calls to $\mathcal{S}_{2k}$, which group into three exponential factors:
\begin{equation}
\mathcal{S}_{2k+2}(\Delta t) = e^{2\mathcal{A}}\,e^{\mathcal{B}}\,e^{2\mathcal{A}},
\end{equation}
where, using $[\mathcal{S}_{2k}(\tau)]^2 = e^{2\ln \mathcal{S}_{2k}(\tau)}$ and Eq.~\eqref{eq:inductive_hypothesis},
\begin{align}
2\mathcal{A} &= -i\bigl(2p_k H\,\Delta t + 2p_k^{2k+1}C_{2k+1}\,\Delta t^{2k+1}+O(\Delta t^{2k+3})\bigr), \\
\mathcal{B}  &= -i\bigl(s_k H\,\Delta t + s_k^{2k+1}C_{2k+1}\,\Delta t^{2k+1}+O(\Delta t^{2k+3})\bigr).
\end{align}
At leading order ($O(\Delta t)$), all three factors are proportional to the same operator $H$.

To show that the $O(\Delta t^{2k+1})$ term in $\ln \mathcal{S}_{2k+2}$ vanishes, we examine two contributions:

(a)~\emph{Internal $C_{2k+1}$ terms:} These sum over all five $\mathcal{S}_{2k}$ calls with total coefficient
\begin{equation}
4p_k^{2k+1} + s_k^{2k+1}.
\end{equation}
The Suzuki parameter $p_k = 1/(4-4^{1/(2k+1)})$ is chosen precisely to make this vanish.

(b)~\emph{BCH cross-commutators between different factors:} Since previous recursion levels have eliminated all intermediate odd powers $O(\Delta t^3),\ldots,O(\Delta t^{2k-1})$, the only sub-leading building blocks at $O(\Delta t^{2k+1})$ are $(2k{+}1)$-fold nested commutators of the $O(\Delta t)$ terms $\alpha_i H\,\Delta t$:
\begin{equation}
[\alpha_{i_1}H,\,[\alpha_{i_2}H,\,[\cdots[\alpha_{i_{2k}}H,\,\alpha_{i_{2k+1}}H]\cdots]]] \propto \underbrace{[H,[H,\cdots[H,H]\cdots]]}_{2k+1} = 0,
\end{equation}
which vanish by $[H,H]=0$. One might also consider cross-terms that mix an internal $C_{2k+1}\,\Delta t^{2k+1}$ piece from one factor with the $O(\Delta t)$ leading term of another factor via a BCH commutator, e.g.\ $[\alpha_i H\,\Delta t,\,C_{2k+1}\,\Delta t^{2k+1}] = O(\Delta t^{2k+2})$. Such terms are beyond the order being analyzed and do not affect the cancellation at $O(\Delta t^{2k+1})$.

Combining (a) and (b), the $O(\Delta t^{2k+1})$ term in $\ln \mathcal{S}_{2k+2}$ vanishes. The leading error is therefore pushed to $O(\Delta t^{2k+3})$. Since $\mathcal{S}_{2k+2}$ is built from symmetric factors, the identity $\mathcal{S}_{2k+2}(-\Delta t)=\mathcal{S}_{2k+2}(\Delta t)^{-1}$ holds, so its generator again contains only odd powers of $\Delta t$, so dividing by $-i\Delta t$ gives
\begin{equation}
H_{\mathrm{eff}}^{(2k+2)} = H + \Delta t^{2k+2}K_{2k+2}+O(\Delta t^{2k+4}),
\end{equation}
where $K_{2k+2}$ is a Hermitian operator determined by the $O(\Delta t^{2k+3})$ term in $\ln\mathcal{S}_{2k+2}$. This completes the induction. \hfill$\square$

For any $2k$-order formula, the perturbation $V=\Delta t^{2k}K_{2k}$ enters the interaction-picture formalism exactly as in the second-order derivation, with the substitution $\Delta t^2 K_2 \to \Delta t^{2k}K_{2k}$.
The error state becomes
\begin{equation}
|\delta\psi(t)\rangle \approx i\,\Delta t^{2k}\int_0^t d\tau\,e^{-iH(t-\tau)}\,K_{2k}\,e^{-iH\tau}\,|\psi_0\rangle,
\end{equation}
and expanding in the eigenbasis of $H$ yields
\begin{equation}
\|\delta\psi(t)\|^2 \approx \Delta t^{4k}\sum_n\left|\sum_m c_m(K_{2k})_{nm}\,e^{\frac{i\omega_{nm}t}{2}}\frac{2\sin\frac{\omega_{nm}t}{2}}{\omega_{nm}}\right|^2.
\label{eq:error_general}
\end{equation}
The stroboscopic factor $\sin(\omega_{nm}t/2)$ appears identically for every order $2k$, depending only on the eigenspectrum of $H$. This completes the proof that the error suppression mechanism, and hence the Trotter-scar phenomenon, extends to $(2k)$th-order Suzuki formulas of arbitrary order.

\section{First-order Lie-Trotter decomposition}
\label{app:first_order}

The preceding sections focused on even-order Suzuki formulas ($q=2,4,\ldots,2k$).
A natural question is whether the Trotter scar mechanism also operates for the simplest product formula, the first-order Lie-Trotter decomposition.
We show here that it does: the periodic vanishing of the stroboscopic factor is equally present for $q=1$.

\subsubsection{Definition and BCH expansion}

The first-order (Lie--Trotter) decomposition is
\begin{equation}
\mathcal{S}_1(\Delta t) = e^{-iH_o\,\Delta t}\,e^{-iH_e\,\Delta t}.
\label{eq:lie_trotter}
\end{equation}
Unlike the second-order symmetric decomposition $\mathcal{S}_2$, this formula does \emph{not} possess the symmetric splitting structure.
To see this explicitly, note that
\begin{equation}
\mathcal{S}_1(-\Delta t) = e^{iH_o\,\Delta t}\,e^{iH_e\,\Delta t},
\end{equation}
whereas
\begin{equation}
\mathcal{S}_1(\Delta t)^{-1} = e^{iH_e\,\Delta t}\,e^{iH_o\,\Delta t},
\end{equation}
and these differ whenever $[H_o, H_e]\neq 0$.
Consequently, the identity $\mathcal{S}(-\Delta t)=\mathcal{S}(\Delta t)^{-1}$ that enforces the absence of even powers in $\ln \mathcal{S}_2$ does \emph{not} hold for $\mathcal{S}_1$.

The BCH expansion of $\ln \mathcal{S}_1(\Delta t)$ therefore contains \emph{all} powers of $\Delta t$, both even and odd.
Writing $X=-iH_o\,\Delta t$ and $Y=-iH_e\,\Delta t$, the standard BCH formula gives
\begin{equation}
\ln(e^X e^Y) = X + Y + \tfrac{1}{2}[X,Y] + \tfrac{1}{12}\bigl([X,[X,Y]]+[Y,[Y,X]]\bigr) + \cdots
\end{equation}
Substituting and collecting by powers of $\Delta t$:
\begin{align}
\ln \mathcal{S}_1(\Delta t)
&= -iH\,\Delta t
\underbrace{- \tfrac{1}{2}[H_o,H_e]\,\Delta t^2}_{C_2}
\nonumber\\
&\quad + \underbrace{\tfrac{i}{12}\bigl([H_o,[H_o,H_e]] - [H_e,[H_o,H_e]]\bigr)\,\Delta t^3}_{C_3}
+ O(\Delta t^4),
\label{eq:bch_first_order}
\end{align}
where we used $[X,Y] = (-i)^2[H_o,H_e]\,\Delta t^2 = -[H_o,H_e]\,\Delta t^2$ and similarly for the higher terms.

Since $C_2 = -\tfrac{1}{2}[H_o,H_e] \neq 0$ generically, the leading error appears at $O(\Delta t^2)$, and the formula is first-order: $q=1$.

\subsubsection{Effective Hamiltonian and error kernel}

Dividing by $-i\,\Delta t$, the effective Hamiltonian reads
\begin{equation}
H_{\mathrm{eff}}^{(1)} = H + \Delta t\, K_1 + O(\Delta t^2),
\label{eq:heff_first}
\end{equation}
where the leading error kernel is
\begin{equation}
K_1 = -\frac{i}{2}[H_o, H_e].
\label{eq:K1}
\end{equation}
We verify that $K_1$ is Hermitian: since $H_o$ and $H_e$ are both Hermitian, $[H_o,H_e]^\dagger = [H_e,H_o] = -[H_o,H_e]$, so the commutator is anti-Hermitian.

Since $H_{\mathrm{eff}}^{(1)}$ has the structure $H + \Delta t^q K_q$ with $q=1$, the general framework of the preceding sections applies directly.
Substituting $q=1$ into the interaction-picture perturbation theory, the error state becomes
\begin{equation}
|\delta\psi(t)\rangle \approx i\,\Delta t\int_0^t d\tau\,e^{-iH(t-\tau)}\,K_1\,e^{-iH\tau}\,|\psi_0\rangle.
\end{equation}
The spectral expansion proceeds identically. Expanding in the eigenbasis of $H$ and evaluating the time integral, the squared error norm is
\begin{equation}
\|\delta\psi(t)\|^2 \approx \Delta t^{2}\sum_n\left|\sum_m c_m(K_1)_{nm}\,e^{\frac{i\omega_{nm}t}{2}}\frac{2\sin\frac{\omega_{nm}t}{2}}{\omega_{nm}}\right|^2.
\label{eq:error_first_order}
\end{equation}
The stroboscopic factor $\sin(\omega_{nm}t/2)$ is identical to that appearing in Eq.~(6) of the main text, confirming that the Trotter-scar mechanism operates for the first-order Lie--Trotter decomposition as well.

\section{Spectral support of the optimized states}
\label{app:spectral}

\begin{figure}[h]
    \centering
    \includegraphics[width=\textwidth]{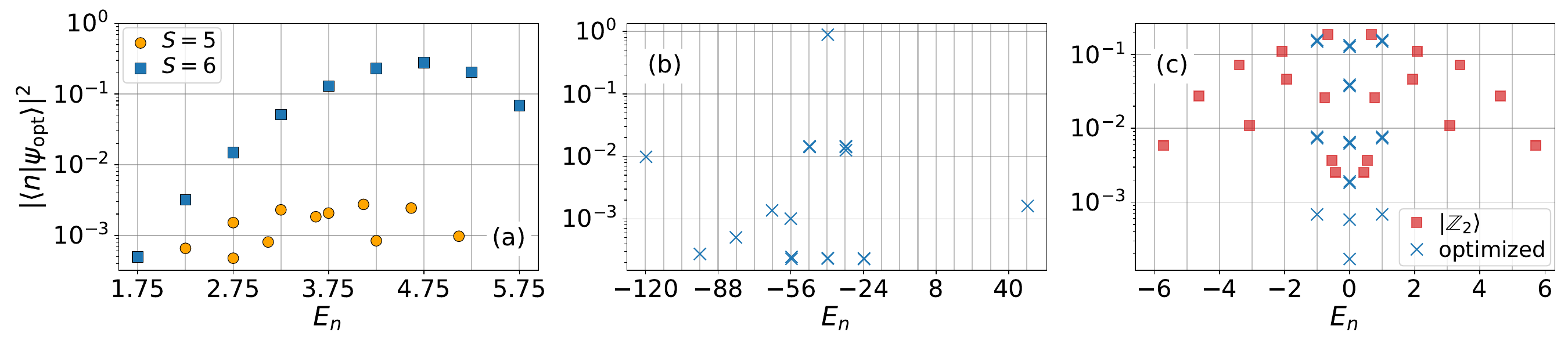}
    \caption{Eigenstate overlap $|\langle n|\psi_{\mathrm{opt}}\rangle|^2$
    versus eigenenergy $E_n$ for the 20 eigenstates with the largest
    overlap for (a)~the Heisenberg chain, (b)~the Stark chain and
    (c)~the PXP model.
    The vertical gray lines mark the commensurate energy grid with
    spacing $\Omega$.
    In~(a) eigenstates are labeled by their total spin quantum number
    $S$ obtained from simultaneous diagonalization of $H$,
    $\mathbf{S}^2$, and $S^x$.
    In~(c) the optimized state is compared with the N\'eel state
    $|\mathbb{Z}_2\rangle$.
    Parameters are the same as in Fig.~1 of the main text.}
    \label{fig:ladders}
\end{figure}

This section presents the spectral decomposition of the optimized
states of the main text.
Figure~\ref{fig:ladders} shows the eigenstate overlap
$|\langle n|\psi_{\mathrm{opt}}\rangle|^2$ versus the eigenenergy
$E_n$.
Each panel shows the 20 eigenstates with the largest overlap.
In all three models the spectral weight concentrates on equally spaced
energy levels.
This confirms the commensurability condition $\omega_{nm}=k\,\Omega$
predicted by Eq.~(6) of the main text.
In the Heisenberg chain the dominant weight lies in a single spin
multiplet split by the transverse field.
In the Stark chain the supporting levels follow the ladder of the
strong Stark field with spacing $\Omega=2h_z$.
In the PXP model we additionally show the spectral weight of the
N\'eel state $|\mathbb{Z}_2\rangle$.
Its weight spreads over the scar tower whose spacing is only
approximately commensurate.
The optimized state instead concentrates its weight on the
integer-spaced levels of the blockaded subspace.

\section{Device parameters}
\label{sec:device}

The experiments in Fig.~1(g--i) of the main text were performed on a
17-qubit superconducting processor accessed through the Logical Qubit
quantum cloud platform.
The processor is offered in two lattice configurations.
The repetition-code configuration connects the 17 qubits into a linear
chain with 16 couplers.
The surface-code configuration connects the same qubits in a
two-dimensional bipartite layout with 24 couplers.
The native gate set consists of arbitrary single-qubit rotations and the
CZ gate.
Our circuits run on nearest-neighbor subchains embedded in these
layouts.
The Heisenberg and Stark $L=4$ data were taken in the surface-code
configuration and all other data in the repetition-code configuration.
Table~\ref{tab:device} summarizes the calibration metrics reported by
the platform at the time of the experiments.

\begin{table}[h]
\caption{Calibration metrics of the two lattice configurations of the
17-qubit processor.
The platform identifiers are \texttt{QZ01-repetition\_code} and
\texttt{QZ01-surface\_code}.
Values are medians over the device and brackets give the full range.
The single-qubit gate fidelity is the reported XEB fidelity.
The readout fidelities $F_{00}$ and $F_{11}$ are the probabilities of
correctly reading out a qubit prepared in $\ket{0}$ and in $\ket{1}$.
The snapshot dates are 2026-07-23 for the repetition-code layout and
2026-07-24 for the surface-code layout.}
\label{tab:device}
\begin{ruledtabular}
\begin{tabular}{lcc}
 & Repetition-code layout & Surface-code layout \\
\hline
Qubits & 17 & 17 \\
Couplers & 16 (linear chain) & 24 (bipartite) \\
$T_1$ ($\mu$s) & 34.7 [17.0, 61.9] & 37.3 [16.7, 54.1] \\
$T_2^{\mathrm{E}}$ ($\mu$s) & 9.0 [7.1, 14.8] & 8.6 [7.1, 15.0] \\
$T_2^{*}$ ($\mu$s) & 2.8 [1.6, 6.2] & 2.7 [1.6, 6.3] \\
Single-qubit gate fidelity & 0.9993 [0.9988, 0.9995] & 0.9993 [0.9992, 0.9995] \\
CZ gate fidelity & 0.9950 [0.9916, 0.9963] & 0.9950 [0.9914, 0.9965] \\
Readout fidelity $F_{00}$ & 0.999 [0.996, 1.000] & 0.999 [0.995, 1.000] \\
Readout fidelity $F_{11}$ & 0.979 [0.968, 0.986] & 0.979 [0.972, 0.983] \\
\end{tabular}
\end{ruledtabular}
\end{table}

\end{document}